\documentclass[draft,showpacs,preprintnumbers,amsmath,amssymb]{revtex4-2}
\usepackage[utf8]{inputenc} 
\usepackage{soul}
\usepackage{xcolor}
\usepackage{array}
\usepackage{caption}
\usepackage{subcaption}
\usepackage{graphicx}
\setstcolor{red}
\begin{document}

\title{Straight and Wiggly Cosmic Strings in  Horndeski Theory}

\author{M. Haluk Se\c cuk}
\email{haluk.secuk@marmara.edu.tr}
\affiliation{Physics Department, Marmara University, Faculty of Science, \.Istanbul, T\"urkiye}

\author{\"Ozg\"ur Delice}
\email{ozgur.delice@marmara.edu.tr}
\affiliation{Physics Department, Marmara University, Faculty of Science, \.Istanbul, T\"urkiye}

\date{\today}
\begin{abstract}

In this article, the behavior of a straight cosmic string is studied for the linearized version of  Horndeski theory in cases where the scalar field is massless or massive. Several physical properties of such solutions are discussed in detail regarding the effects of the scalar field of this theory. The mass of the scalar field induces a screening effect such that, in the massive theory, the string solution approaches to the general relativistic one. We also consider wiggly cosmic strings, obtain the solutions for both massless and massive cases, discuss their properties and observe similar screening effects. 
\end{abstract}
\maketitle

\section{Introduction}

Cosmic strings \cite{Vilenkin,BariolaVilenkin,VilenkinShellard} are linear topological defects which may be produced  \cite{Kibble}  during symmetry-breaking phase transitions in the early universe. The very peculiar properties of these exotic objects make them appealing to study their properties and  deduce their cosmological and astrophysical consequences.  When they formed in the early Universe, a cosmic string network would have been composed of either infinite strings or string loops \cite{Kibble,Albrecht}. Infinite strings may produce wakes in the cosmic background when they move \cite{wake}. It is thought that the density fluctuations leading to structure formation in the early Universe may have originated by the evolution of cosmic string networks. The aforementioned network emits electromagnetic, gravitational, and other forms of radiation when it evolves and relaxes. This evolution is rather complex and model-dependent, requiring numerical simulations to understand it. However, data accumulated at the end of the last century revealed that the primary source of density fluctuations has a quantum origin. These observations may reduce the role played by topological defects  in structure formation in the early universe. Although the contribution of topological defects, including cosmic strings, to structure formation is not completely ruled out, it has been shown that it is very limited \cite{Pogosian,Pogosian1,Bevis}, if these objects really exist.

The gravitational properties of cosmic strings are also a subject of interest. For example, in general relativity (GR), the gravitational potential of an infinitely long straight cosmic string vanishes. Therefore, the spacetime around such a string is locally flat. It is globally conical, however, since the line element involves an angle deficit proportional to the tension of the string. This global effect produces observable astrophysical effects, such as double images of galaxies or other distant objects, production of imprints, for instance, line discontinuities and anisotropies on the cosmic microwave background radiation (CMB). Although cosmic strings have not been observed yet, they still await their discovery. The astrophysical, cosmological,  and gravitational wave observations set limits on the physical parameters of the hypothetical cosmic strings; one such parameter is the string tension. 


Numerical and theoretical investigations regarding cosmic string networks imply that small-scale perturbations such as wiggles and kinks may be realized in the cosmic string network \cite{Vilenkin1,Vachaspati}. For a distant observer who cannot resolve these perturbations on a long string, the string will appear smooth \cite{Vachaspati,VilenkinShellard}. However, the effect of these perturbations will be realized in the mass per unit length and tension of the string \cite{Vilenkin1}. These changes will have consequences on the gravitational and cosmological properties of wiggly strings \cite{VilenkinShellard}, which makes their further properties worth investigating \cite{Vachaspati1,Ozdemir,Azevedo}. 

Theories alternative to or modifying GR are still an active research topic for various observational and theoretical reasons. We refer to the excellent reviews \cite{Capozziello,Clifton,Joyce,Nojiri,Shankaranarayanan} on the modified gravity theories for these motivations and recent developments. One way to distinguish such theories is to consider the same physical setup for the alternative theories, identify their outcomes, and then use observations to rule out or constrain the theories that conflict with the observations. Here, we will study cosmic strings and obtain their various properties for the theory we consider, and compare them with GR. There are various studies considering straight \cite{GundlachOrtiz,BarrosRomero,Guimaraes,Sen,Sen1,Delice, Delice1, Harko, Harko1,Harko2,Momeni} and wiggly \cite{AraziwgBD} cosmic strings in alternative theories in the last few decades.  

In this article, we would like to investigate the gravitational field of a straight or wiggly cosmic string in the Horndeski theory \cite{Horndeski}. This theory is  the  most general scalar-tensor theory, leading to second-order field equations in four dimensions \cite{Horndeski,Deffayet}. We can say that it acts as a general framework theory, whose special cases include GR, Brans-Dicke theory (BD),  $f(R)$ theory, etc. Hence, most scalar-tensor theories are special cases of the Horndeski theory \cite{KobayashiReview}. The field equations of this theory are very complex \cite{Kobayashi}; hence, we keep our discussion at the linearized level. Even at this level, several phenomenological questions can be answered, such as the behavior of the spacetime near the core of the string and also the behavior of test particles around these cosmic strings.

The article is organized as follows. In section two, we present the weak-field equations of Horndeski theory with a massless or massive scalar field for a straight cosmic string. We obtain their solutions in section three for a massless or massive scalar field. The section IV will be devoted to obtaining several physical implications of these solutions. We discuss the solutions of wiggly cosmic strings in the fifth section and their properties in the sixth section. Note that the solutions we will obtain for massless Horndeski theory resemble the form of the corresponding solutions  for BD theory for both straight \cite{GundlachOrtiz,BarrosRomero} and wiggly \cite{AraziwgBD} strings. Whereas corresponding solutions for massive Horndeski theory are, as far as we know, considered for the first time here. And finally, the seventh section is devoted to studying the unbounded motion of particles passing through these strings. We conclude this paper with a brief discussion.

\section{Horndeski theory and Linearized field equations}\label{sec1}

The Lagrangian of the action $S=\int\mathcal{L} \sqrt{|g|}d^4x$ of Horndeski theory \cite{Horndeski}, also known as the generalized Galileon \cite{Deffayet}, can be written as \cite{KobayashiReview}:
\begin{eqnarray}
\mathcal{L}&=&K(\phi,X)-G_3(\phi,X)\square \phi  + G_4(\phi,X) R  \nonumber \\ && + G_{4,X}\left[ \left( \square \phi\right)^2- (\nabla^\mu \nabla^\nu\phi) (\nabla_\mu \nabla_\nu\phi) \right] + G_5(\phi,X) G_{\mu\nu}\nabla^\mu \nabla^\nu \phi  \nonumber \\ &&  -\frac{G_{5,X}}{6} \left[\left(\square \phi \right)^3-3 \square\phi\, (\nabla^\mu \nabla^\nu\phi) (\nabla_\mu \nabla_\nu\phi)+2 (\nabla_\mu \nabla^\nu\phi)(\nabla_\nu \nabla^\lambda\phi)(\nabla_\lambda \nabla^\mu\phi) \right], \label{Horndeskiaction}
\end{eqnarray}
where $X=-\nabla^\mu\nabla_\mu \phi /2$ is a kinetic term and $K,G_3,G_4, G_5$ are arbitrary functions of the scalar field $\phi$ and kinetic term $X$. We use the notation $f_{,X}=\partial f/\partial X$. We can add a minimally coupled matter Lagrangian $\mathcal{L}_m$ to  (\ref{Horndeskiaction}) to explore the behavior of the matter-energy distribution in this theory.
Since the  field equations of this theory are rather complicated, we refer to \cite{Kobayashi}  for them.

In this work, we consider linearized field equations of Horndeski theory. We will present the details  of these equations in the Appendix \ref{Appendix} and here only present the relevant equations to obtain the solutions for clarity.

We consider a weak field expansion of the metric tensor and the scalar field as
\begin{eqnarray}
&&	g_{\mu\nu}=\eta_{\mu\nu} + h_{\mu\nu},\\
&&	\phi=\phi_0+ \varphi,	
\end{eqnarray}
where $\eta_{\mu\nu}$ is the Minkowski metric,  $h_{\mu\nu}$ is the metric perturbation tensor and  $\phi_0$ a constant with the conditions $h_{\mu\nu}\ll 1$, $\varphi\ll 1$. It turns out that it  is better to express linearized field equations in terms of an auxiliary tensor  $\theta_{\mu\nu}$ defined in (\ref{thetadef})   whose relation with   the metric perturbation tensor is given as 
\begin{equation}
 h_{\mu\nu}=\theta_{\mu\nu}-\frac{1}{2}\eta_{\mu\nu}\,\theta -\eta_{\mu\nu}\frac{G_{4,\phi}(0)}{G_4(0)}\varphi. 
\end{equation}
Together with Lorenz gauge condition given in (\ref{Lorenz}), the weak field equations we need to solve reduce to
\begin{eqnarray}
	\square_\eta \theta_{\mu\nu}&=&-\frac{ 2\kappa}{G_4(0)}\, T_{\mu\nu}^{(1)}, \label{tensoreq} \\
		(\square_\eta -m^2)\varphi &=& \kappa'\, T^{(1)}.\label{scalareq}
\end{eqnarray} 
Here $\square_\eta$ is the  D'Alembertian operator of the flat Minkowski spacetime, $m$ is the mass of the scalar field,  defined as (\ref{massterm}), and the coupling constant $\kappa'$ is defined as (\ref{kappaprime}).   Now,  we can discuss the solutions corresponding to straight  cosmic strings in the following section.

\section{Solutions for straight Cosmic Strings}
The energy momentum tensor for a straight cosmic string lying along the $z$ axis, in the zero thickness approximation, is given by \cite{Vilenkin1}
\begin{equation}
T_{\mu \nu}=\mu\ \delta(x)\delta(y) \mbox{diag}(1,0,0,-1),
\end{equation}
in Cartesian coordinates with $x^\mu=(t,x,y,z)$, where $\mu$ is the tension of the string. For this energy-momentum tensor, the nontrivial Horndeski field equations (\ref{tensoreq}) read
\begin{eqnarray}
\square \theta_{00}=-\square \theta_{zz}=\frac{-2\kappa}{G_4(0)} \mu\, \delta(x)\delta(y),
\end{eqnarray}
with solutions
\begin{equation}\label{Thetasolutions}
	\theta_{00}=-\theta_{zz}=-\frac{\kappa\, \mu}{\pi G_4(0)}  \ln \frac{r}{r_0},\quad \theta_{xx}=\theta_{yy}=0.
\end{equation}
Here $r=\sqrt{x^2+y^2}$ and $r_0$ is an integration constant.
The scalar field equation, (\ref{scalareq}), now becomes
\begin{equation}
	(\square_\eta -m^2)\varphi=-2\kappa'\,\mu \, \delta(x)\delta(y).\label{scalarcseq}
\end{equation}
The solutions of this equation will be different for massless and massive scalar field cases. Hence, we obtain them separately in the following.

\subsection{Massless scalar field case}
When the function $K$ in the action (\ref{Horndeskiaction}) is at most linear in the scalar field $\phi$, the mass of the scalar field $m$ defined in (\ref{massterm}) vanishes. Therefore, the scalar field equation (\ref{scalarcseq}) has the following solution for massless case, i.e. $m=0$,

\begin{equation}
	\varphi=-\frac{\kappa'\,  \mu}{\pi }\,\ln \frac{r}{r_0}.
\end{equation}
Using this solution, (\ref{thetadef}) and (\ref{Thetasolutions}), the metric perturbation tensor has nontrivial components given by
\begin{eqnarray}
	&&h_{00}=-h_{zz}=-\frac{G_{4,\phi}(0)}{\pi G_4(0)}\kappa'\mu\, \ln \frac{r}{r_0},
    \\ &&h_{xx}=h_{yy}=\frac{G_{4,\phi}(0)\kappa'-\kappa}{\pi G_4(0)}\,\mu\, \ln \frac{r}{r_0}.
\end{eqnarray}
Hence, employing the cylindrical coordinates, the spacetime metric reads
\begin{eqnarray}
ds^2=\left[1+\frac{G_{4,\phi}(0)}{\pi G_4(0)}\kappa'\mu\ln \frac{r}{r_0} \right]
\left\{-dt^2+dz^2+\left[1-\frac{\kappa\, \mu}{\pi G_4(0)} \ln \frac{r}{r_0}\right] \left(dr^2+r^2 d\theta^2 \right) \right\}. 
\end{eqnarray}	
 The transformation $r=r_0\,(r'/a)^b$ with $a=[1-\kappa/(2\pi G_4(0))\mu]^{-1/2} $ and $b=[1-\kappa/(4\pi G_4(0))\mu]^{-1}$, adapted from \cite{BarrosRomero}, brings the line element into a more familiar form
\begin{eqnarray}\label{csmassless}
	ds^2=\left[1+\frac{G_{4,\phi}(0)}{\pi G_4(0)}\kappa'\mu\ln \frac{r'}{r_0}\right]
    \left\{-dt^2+dz^2+dr'^2+\left[ 1-\frac{\kappa\, \mu}{\pi G_4(0)}  \right] r'^2 d\theta^2 \right\}.
\end{eqnarray}	
Hereafter we will set $r_0=1$ without loss of generality. Under the GR limit $\kappa'=0,G_{4}(0)=1$, $G_3=G_5=K=0$, this solution reduces to Vilenkin's weak field solution \cite{Vilenkin1}, whereas in the BD limit, i.e. $G_4=\phi$, $G_3=G_5=0$, $K=2\omega X/\phi$, this solution reduces to weak field solution presented in \cite{GundlachOrtiz,BarrosRomero} which is known to be conformal to Vilenkin's solution. When the conformal factor survives, the solution is valid only near the string, since the weak field approximation breaks down towards radial infinity, as in the BD case \cite{GundlachOrtiz,BarrosRomero}. Apart from GR, however, for any subclass of Horndeski theory where $G_{4,\phi}(0)=0$ and the conformal factor vanishes, including Einstein-scalar theory, Vilenkin's weak field solution is also a solution for any value of radial coordinate, at least in the linearized regime. Whether the mass of the scalar field remedies this behavior will be the subject of the following subsection.

\subsection{Massive scalar field case}

\begin{figure}[t] 
\includegraphics[draft=false,width=.7\linewidth]{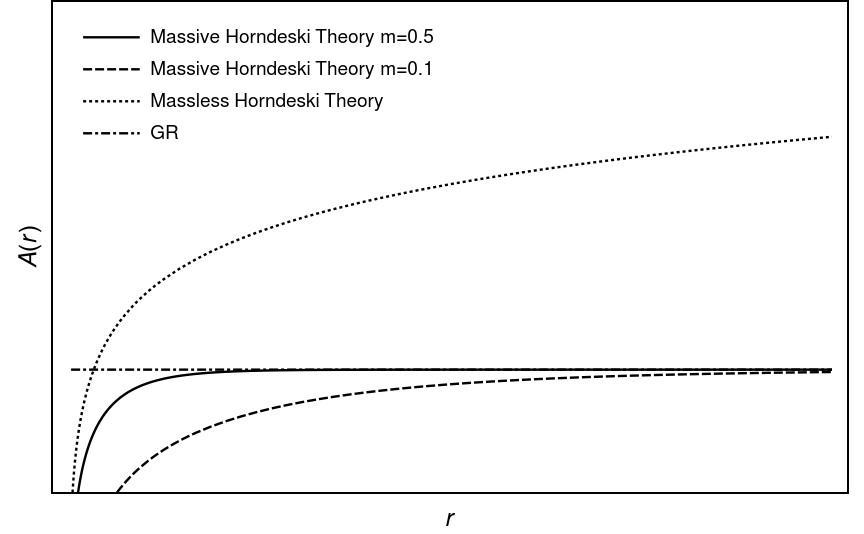}
		\caption{Behavior of the conformal factors  of massless, massive with $m=0.1$, $m=0.5$    and general relativity cases. It is clear that in the massive case, the conformal factor approaches to unity, i.e. the  GR limit for large $r$ values, whereas in the massless case it increases by increasing $r$. The general behavior of massive and massless cases agrees with small values of $r$ as the mass of the scalar field decreases. Note that plots are not scaled; We choose constant values such that these behaviors are clear to see on a single graph.\raggedright}
	\label{fig:1}
\end{figure}

In this case, the general solution of the scalar field equation (\ref{scalarcseq}) outside the string core becomes
\begin{equation}
    \varphi(r)=c_0\, I_0(m r)+c_1\, K_0( m r)-\frac{\kappa' \mu}{\pi}\, K_0(m r),
\end{equation}
where $I_\alpha$ and $K_\alpha$ are  modified Bessel functions of the first and second kind, and $c_0$ and $c_1$ are arbitrary integration constants, and they should be in the same order as $\kappa'\mu$. In order to fix the integration constants $c_0$ and $c_1$, we choose the following boundary conditions: (i) we require  the scalar field to vanish at the radial infinity, $\varphi(r\rightarrow \infty )\approx 0$, (ii) near the string, the solution to resemble similar behavior with the massless  case. Then, for (i) we can choose $c_0=0$ since $I_0$ blows up but $K_0$ vanishes at infinity. This is in contrast to the corresponding solution in the massless case. The boundary condition (ii) can be satisfied if we choose, without loss of generality, $c_1=2\kappa'\mu/\pi$. Note that with this choice, the scalar field of the massless case and the massive case near the string have similar behavior.

We will show that this choice is justified when we discuss the conformal factors of both massless and massive cases in the following subsection. Under these assumptions, the massive scalar field becomes
\begin{equation}\label{strmasspot}
	\varphi(r)=\frac{\kappa' \mu}{\pi}\, K_0(m r). 
\end{equation}

For the massive scalar field case, we obtain the following line element
\begin{eqnarray}
ds^2=&&\left[1-\frac{G_{4,\phi}(0)}{\pi\,G_4(0)}\kappa'\mu \, K_0(m r) \right]
\left\{-dt^2+dz^2+\left[1-\frac{\kappa \mu }{ \pi\, G_4(0)}  \ln r \right]\left(dr^2+r^2 d\theta^2\right) \right\}.
\end{eqnarray}
We can also express this line element  in the transformed coordinates as  given in \cite{Vilenkin1}, using again \cite{BarrosRomero} as in previous subsection,  it becomes  linear in $\mu$  as 
\begin{eqnarray}\label{csmassive}
	ds^2=\left[1-\frac{ G_{4,\phi}(0)}{\pi G_4(0)}\kappa'\mu\, K_0( m r') \right]
    \left\{-dt^2+dz^2+dr'^2+\left[ 1-\frac{\kappa\,\mu }{\pi G_4(0)}  \right] r'^2 d\theta^2 \right\}.	
\end{eqnarray}	
For  the large values of  $r'$,  the function $K_0$ diverges towards zero, hence  the line element becomes
\begin{equation}\label{csmassiveasy}
	ds^2_{r\rightarrow \infty}\approx -dt^2+dz^2+dr'^2+\left[ 1-\frac{\kappa\, \mu }{\pi G_4(0)}  \right] r'^2 d\theta^2,
\end{equation}	
which clearly approaches Vilenkin's solution in GR. Therefore, we see that the general behavior of cosmic strings in massive Horndeski theory differs from the massless case. In some sense, the mass of the scalar field screens the effects of the scalar field, leading to the GR result far from the string. It is well known that massive theories show this behavior \cite{Vainshtein,Khoury,Khoury1}, where the scalar field freezes beyond a point. For distant regions from the string, we have a locally flat spacetime with an angle deficit, whereas the behavior of particles near the string will be altered by the conical space behavior of GR cosmic strings. However, the behavior of photons will be the same with GR since the line elements are conformal to the GR solution.
 \begin{figure}[t] 
\includegraphics[draft=false,width=.7\linewidth]{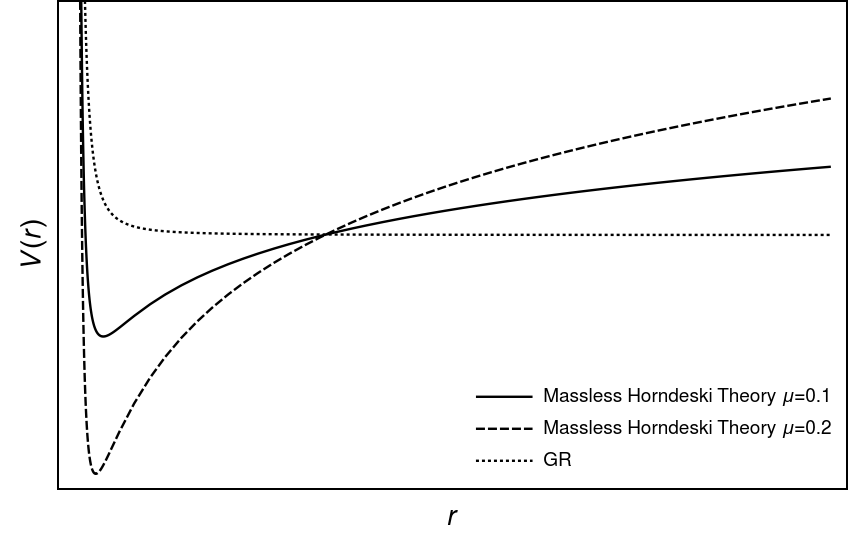}
	\caption{Behavior of the effective potential of massless Horndeski theory with $\mu=0.1$ and $\mu=0.2$. Observe that, when we double the string tension $\mu$, the trapping energy of a particle decreases and the depth of the trapping region increases. Meaning, a particle that is trapped in a gravitational region with a higher string tension in massless Horndeski theory has a low chance of escaping from that field. The graph is plotted for $L=0.01$. Other constants are chosen such that the behavior is reflected on a single graph.\raggedright
    }
    \label{fig:2}
\end{figure}
\section{Properties of Horndeski Cosmic Strings}
\subsection{Conformal Factor}
The solutions of Horndeski cosmic strings are conformal to those of GR. The conformal factor determines the character and differences of these objects in the corresponding theories. In the case of the massless Horndeski theory, the conformal factor is an increasing function of $r$; as a result, the solution is valid only near the string where the linear approximation is valid. However, for the far region, one needs to look for exact solutions to describe the behavior of these strings. By contrast, in the massive Horndeski theory, the conformal factor asymptotically approaches unity, i.e., the GR solution, which makes the solution valid everywhere. These distinct behaviors can be clearly seen in Figure (\ref{fig:1}). 

\subsection{Angular Deficit}

It is well known that, for straight cosmic strings in GR, the spacetime geometry is locally pseudo-Euclidean but globally conical due to the angular deficit parameter in the line element. This can be seen if we define a new angular coordinate $\theta'=(1-\kappa\mu/2\pi)\,\theta$, then the line element of a straight cosmic string in GR becomes $ds^2=-dt^2+dr^2+dz^2+r^2 d\theta'^2$, but the range of the angular coordinate is now $\theta'\in [0,2\pi (1-\kappa\mu/2\pi)]$, which clearly defines a conical spacetime. Since the spacetime is locally Euclidean, a cosmic string does not apply any gravitational attraction to test particles or photons around it, and their trajectories are straight lines. Hence, the differences in the motion of test particles in this spacetime compared to flat Minkowski
 spacetime have a topological origin. One important outcome of this topological effect is the deflection of light due to a cosmic string. In a light scattering experiment where a photon comes from a very far radial distance and reaches very close to the string and again reaches a very far distance to the string, the change in the angular coordinate is $\Delta\theta'= (1-\kappa\mu/2\pi)\Delta\theta=\pi$ which yields $\Delta\theta=(1+\kappa\mu/2\pi)\pi$, in which we have used the smallness of the angular deficit parameter to derive this result. Therefore, the light deflection angle due to a straight cosmic string in GR is given by
 \begin{equation}\label{defanggr}
   \delta\theta_{GR}=\kappa\mu/2=4\pi G \mu,  
 \end{equation}
which is independent of the closest approach distance to the string.

For the massless Horndeski theory, the line element (\ref{csmassless}) has an overall conformal factor, which means the spacetime is no longer locally flat but conformally locally flat. This implies that the massive test particles will feel an attractive force towards the string, and their trajectories are no longer straight lines. The angular deficit parameter is also no longer constant but increases with increasing radial distance to the string. For the photons, since the  conformal factor does not affect null geodesics, they are still straight rays, and the  deflection angle has the same form as
 \begin{equation}\label{defanghor}
   \delta\theta_{Horndeski}=\frac{\kappa\,\mu}{2\, G_{4}(0)}.
 \end{equation}

For the massive Horndeski theory, we again have a conformal factor in the line element (\ref{csmassive}), and the angular deficit parameter is again a function of radial coordinate. Similar to the massless theory case, the trajectories of massive test particles are clearly affected by this factor. Photon trajectories are not affected by this conformal factor, and the deflection angle for the massive Horndeski theory is  the same as the massless one given in (\ref{defanghor}). One important difference between massless and massive Horndeski theory is that, for the massive theory, the line element becomes asymptotically locally flat as given in (\ref{csmassiveasy}). Hence, for the particles very far from the string, unlike in the massless case, the trajectories are straight lines, and particles feel no force due to the string. Further discussion of particle motion will be presented below for both cases.

\subsection{Geodesics and Effective Potential\label{section4c}}

Let us discuss the motion of test particles and photons around the straight cosmic strings in Horndeski theory. Since the theory we employ is minimally coupled to matter, test particles follow geodesics. The line element that we have derived has the general form
\begin{equation}\label{cshornmetric}
    ds^2=A(r)\left[ -dt^2+dr^2+dz^2+(1-\alpha)\,r^2 d\theta^2\right],
\end{equation}
where $A(r)$ is the conformal factor in the solutions (\ref{csmassless}) or (\ref{csmassive}) and 
\begin{equation}\label{alpha}
\alpha=\frac{\kappa \mu}{\pi\, G_{4}(0)}
\end{equation}
is the angular deficit parameter. We can use the Lagrangian $2L=g_{\mu\nu} \dot x^\mu \dot x^\nu=\epsilon$, where overdot represents derivative with respect to an affine parameter, $\lambda$, and $\epsilon=-1,0$ for particles or photons, respectively.  We find the first integrals of the motion as:
\begin{equation}
	\dot t=\frac{E}{A}  ,\quad \dot \theta=\frac{ L}{A (1-\alpha) r^2}, 
\end{equation}
where $E$ and $L$ are the specific energy and angular momentum of the particle.
Considering the motion on the plane perpendicular to the string with $z=\text{constant}$, we find the following expression 
\begin{equation}
	\dot{r}^2=\frac{1}{A^2}\left[E^2-\frac{L^2}{(1-\alpha)r^2}+\epsilon A\right].
\end{equation}

Considering a new affine parameter $d\lambda=A\, d\gamma$, we obtain an effective potential as
\begin{equation}
	\left(\frac{dr}{d\gamma}\right)^2=E^2-V(r),
\end{equation}
with
\begin{equation}\label{effpot}
V(r)=\frac{L^2}{(1-\alpha)r^2}-\epsilon A(r).
\end{equation}	

Investigating the effective potential, $V$, we see that for massive particles, i. e. $\epsilon=-1$, we see that for the cosmic strings of massless Horndeski theory, there is a potential well and test particles are making bounded motion among turning points given by the equation
\begin{equation}
    E^2=\frac{L^2}{(1-\alpha)r^2} + A,
\end{equation}

\begin{figure}[t]
  \begin{subfigure}[t]{0.5\linewidth}
    \centering
    \includegraphics[draft=false,width=.91\linewidth]{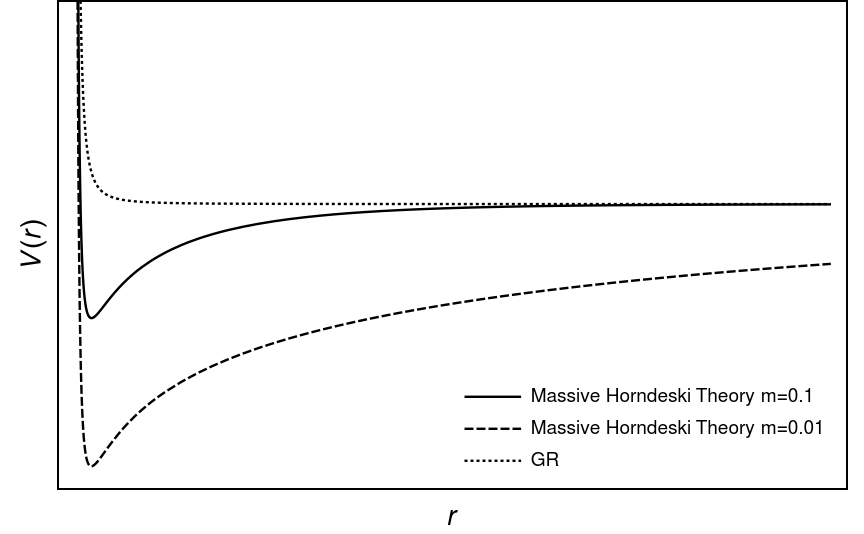} 
    \caption{Behavior of the effective potential of massive Horndeski theory with $m=0.1$ and $m=0.01$. It is clear that there exists a global minimum and then increases by increasing $r$. Hence, we have circular stable geodesics and test particles are trapped in this potential. We see that as the mass of the scalar field decreases, the trapping energy of a particle also decreases.}
    \label{fig3:a} 
    \vspace{2ex}
  \end{subfigure}
  \begin{subfigure}[t]{0.5\linewidth}
    \centering
    \includegraphics[draft=false,width=.91\linewidth]{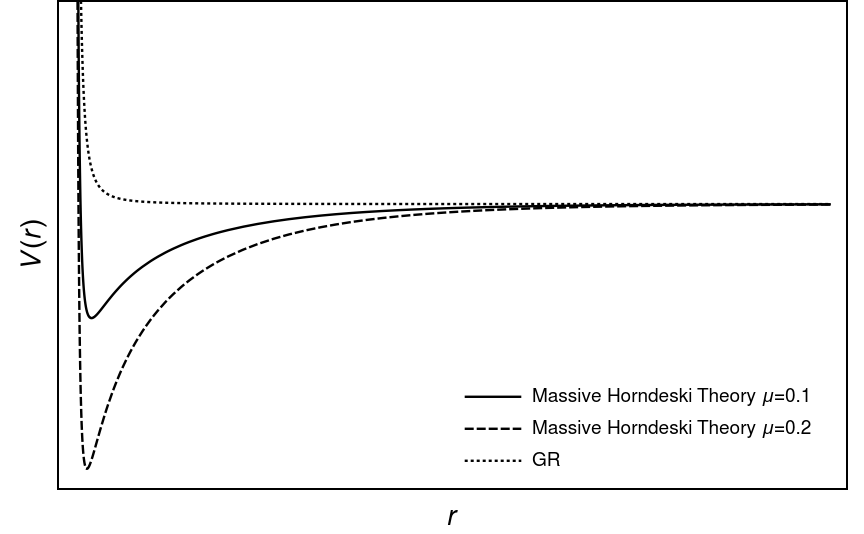} 
    \caption{Behavior of the effective potential of massive Horndeski theory with $\mu=0.1$ and $\mu=0.2$. Observe that, when we double the string tension $\mu$, the trapping energy of a particle decreases and the depth of the trapping region increases. Meaning, a particle that is trapped in a gravitational region with a higher string tension in massive Horndeski theory has a low chance of escaping from that field.} 
    \label{fig3:b} 
  \end{subfigure}
    \caption{Behavior of the effective potential for massive Horndeski theory. Both graphs are plotted for $L=0.1$ and have the same range. In (\ref{fig3:a}), $\mu =0.1$ and in (\ref{fig3:b}), $m=0.2$. Note that plots are not scaled; We choose constant values such that these behaviors are clear to see on a single graph.\raggedright
    }\label{fig:3}
\end{figure} 
as shown in Figure (\ref{fig:2}). For the massive Horndeski theory, in the near region the string is gravitationally active and applies attractive forces to the test particles around it, in contrast to the asymptotic region where the conformal factor vanishes and the potential approaches to a finite value. This can be seen in Figure (\ref{fig:3}). Hence, we have a semi-infinite potential well, depending on the specific energy and angular momentum, and the test particles may or may not escape to radial infinity.  Therefore, there could be bounded motion for particles depending on their specific energy and angular momentum. The particles far from the string (at the  asymptotic region $r \rightarrow \infty$) effectively feel no radial force, since the potential vanishes there, and the case will be similar to local cosmic strings in GR. The figure (\ref{fig:3}) also shows that test particles cannot reach the string for both cases, except for the $L=0$ case.
\subsection{Circular Geodesics}
The shape of the effective potential shows that we can have stable circular geodesics at $r=r_c$ for massive test particles obeying $V(r_c)=E^2$, $V'(r_c)=0$ and $V''(r_c)>0$.
These conditions yield the energy and angular momentum of the test particle following the circular geodesic as
\begin{eqnarray}
    E^2&=&\frac{L^2}{(1-\alpha)r_c^2}+ A(r_c), \label{E}\\
    L^2&=&\frac{1}{2}(1-\alpha ) r_c^3 A'(r_c), \label{L}\\
    V''(r_c)&=&A''(r_c)+\frac{3}{r_c} A'(r_c)>0, \label{Vpp}
\end{eqnarray}
where in the last equation we have used the value of $L^2$ given in (\ref{L}).

For the massless Horndeski theory, the specific energy and angular momentum of the particle explicitly becomes
\begin{eqnarray}
E^2&=&1+\frac{L^2}{(1-\alpha)r_c^2}+ \frac{G_{4,\phi}(0) \kappa'\,\mu}{\pi G_{4}(0)} \, \ln r_c,\\
L^2&=&\frac{1}{2}(1-\alpha)\frac{G_{4,\phi}(0) \kappa'\,\mu}{\pi G_{4}(0)} \, r_c^2,
\end{eqnarray}
and the second derivative of the potential is positive, namely
\begin{equation}
    V''(r_c)=\frac{2}{r_c^2} \frac{G_{4,\phi}(0) \kappa'\,\mu}{\pi G_{4}(0)}\, >0.
\end{equation}
This proves the existence of time-like stable circular geodesics around a straight cosmic string in massless Horndeski theory. 

In the massive Horndeski theory, we have similar results. Namely, investigation of the equations (\ref{E},\ref{L},\ref{Vpp}) shows that there is a stable circular geodesic around a straight cosmic string in the massive Horndeski theory as well. These results show that for both the massless and the massive Horndeski theory, unlike GR, straight cosmic strings are gravitationally active and accumulate matter around them, which may help the  formation of large-scale structures of matter around them, in addition to other mechanisms \cite{VilenkinShellard} known for straight cosmic strings in GR.

\section{Solutions for Wiggly Cosmic strings}
	
The dynamics and properties of wiggly strings are discussed in detail in the excellent monograph of Vilenkin and Shellard \cite{VilenkinShellard}, with the observation that the kinks and wiggles will have small length scales compared to the length scale of a cosmic string network and the string will appear smooth \cite{Vachaspati}. The only observational effect will be that the effective mass per unit length $\tilde \mu$ and the effective tension ($\tilde T$) of the string will obey the equation of state   $\tilde \mu \tilde T=\mu^2$ \cite{Vilenkin1}. For an observer far from the typical length scale of a string, the energy-momentum tensor will effectively have the following form \cite{VilenkinShellard,Vilenkin1}:

\begin{equation}
T_{\mu\nu}=\mbox{diag}(\tilde \mu,0,0,-\tilde T)\,\delta(x)\,\delta(y).
\end{equation}   
For this energy momentum-tensor, the nontrivial linearized field tensor equations (\ref{tensoreq}) will have the following form in cylindrical coordinates 
\begin{eqnarray}
	\square \theta_{00}=\frac{-2\kappa\, \tilde \mu}{G_4(0)} \, \delta(x)\delta(y) ,\quad        \square \theta_{zz}=\frac{2\kappa\, \tilde T}{G_4(0)}  \, \delta(x)\delta(y), \label{thetawcssol}
\end{eqnarray}
with corresponding solutions
\begin{equation}
	\theta_{00}=-\frac{\kappa\,\tilde \mu }{\pi G_4(0)} \, \ln r,\quad \theta_{zz}=\frac{\kappa\, \tilde T }{\pi G_4(0)}  \ln r, \quad \theta_{xx}=\theta_{yy}=0.
\end{equation}
The scalar field equation (\ref{scalareq}) takes the following form for wiggly strings,
\begin{equation}
	(\square_\eta -m^2)\varphi=-\kappa'\, (\tilde \mu+\tilde T) \, \delta(x)\delta(y).\label{scalarwcseq}
\end{equation}
 This equation can give different solutions for massless and massive cases, as in the straight strings. Hence, we consider these cases separately in the following subsections.
\subsection{Massless case} 
For the massless scalar field case, $m=0$, the scalar field equation (\ref{scalarwcseq}) has the solution
\begin{equation}
	\varphi=-\frac{\kappa' (\tilde \mu +\tilde T)}{2\pi }\,\ln r.
\end{equation}
Using this solution and also (\ref{thetadef}) and (\ref{thetawcssol}), the metric perturbation terms become
\begin{eqnarray}
	&&h_{00}=\left[\frac{-\kappa (\tilde\mu-\tilde T)}{2\pi G_4(0)}-\frac{G_{4,\phi} \kappa' (\tilde\mu+\tilde T)}{2\pi G_4(0)}\right]\ln r, \label{wigglyhoomassless}\\
    &&h_{zz}= \left[\frac{-\kappa (\tilde\mu-\tilde T)}{2\pi G_4(0)}+\frac{G_{4,\phi} \kappa' (\tilde\mu+\tilde T)}{2\pi G_4(0)}\right]\ln r,\\
	&&h_{xx}=h_{yy}=\left[\frac{-\kappa (\tilde\mu+\tilde T)}{2\pi G_4(0)}+\frac{G_{4,\phi} \kappa' (\tilde\mu+\tilde T)}{2\pi G_4(0)}\right]\ln r.
\end{eqnarray}

The line element around a linearized wiggly cosmic string in  massless Horndeski theory in cylindrical coordinates  can be written as
\begin{eqnarray}
	ds^2=\left[1+\frac{G_{4,\phi}(0)\kappa' (\tilde\mu+\tilde T)}{2\pi\,G_4(0)}\ln r \right] \Bigg\{&-&\left[1+\frac{\kappa (\tilde\mu-\tilde T)}{2\pi G_4(0)}\ln r\right] dt^2 + \left[1-\frac{\kappa (\tilde\mu-\tilde T)}{2\pi G_4(0)} \ln r \right]dz^2  \nonumber   \\
	&+& \left[1-\frac{\kappa \left( \tilde \mu+\tilde T\right)}{2\pi G_4(0)}\ln r\right] \left(dr^2+r^2 d\theta^2\right) \Bigg\},\label{wigglycsmassless}
\end{eqnarray}

which is conformal to the metric given in \cite{VilenkinShellard}. Therefore, the transformation
\begin{equation}
\left[1-\frac{\kappa}{2\pi G_4(0)}\left( \tilde \mu+\tilde T\right )\ln r\right]r^2=\left[1-\frac{\kappa  \left( \tilde \mu+\tilde T\right) }{2\pi G_4(0)} \right]r'^2 \label{transf1}
\end{equation}
brings the line element in the more familiar form
\begin{eqnarray}
	ds^2=\left[1+\frac{G_{4,\phi}(0)\kappa' (\tilde\mu+\tilde T)}{2\pi\,G_4(0)}\ln r' \right] \Bigg\{&-&\left[1+\frac{\kappa (\tilde\mu-\tilde T)}{2\pi G_4(0)}\ln r' \right] dt^2 + \left[1-\frac{\kappa (\tilde\mu-\tilde T)}{2\pi G_4(0)} \ln r' \right]dz^2  \nonumber   \\
	&+& dr'^2+ \left[1-\frac{\kappa  \left( \tilde \mu+\tilde T\right) }{2\pi G_4(0)} \right] r'^2 d\theta^2 \Bigg\}.
\end{eqnarray}	

This solution is also conformal to the corresponding GR solution given  again in \cite{VilenkinShellard}. We see that the surfaces of constant $t$ and $z$ are conformally equivalent to cones with deficit angle  $\kappa (\tilde \mu +\tilde T)/[2\pi G_4(0)]$. Since the conformal factor is unbounded in radial infinity, this solution can only describe an isolated wiggly string up to the distance where the weak field approximation breaks down. Another important observation is that,  for the appropriate choice of parameters of the Horndeski theory, this solution reduces to the corresponding wiggly string solution for BD theory \cite{AraziwgBD}. Hence, here, we have generalized the wiggly string solution given in \cite{VilenkinShellard} for GR and in \cite{AraziwgBD} into Horndeski theory for the special case where the scalar field is massless. The massive scalar field case will be discussed in the next subsection.

\subsection{Massive scalar field case}
For the case where the scalar field of the Horndeski theory is massive, which originates from the existence of an arbitrary potential in the Horndeski action,  the scalar field equation (\ref{scalarwcseq}) resembles the form of the corresponding straight string equation (\ref{scalarcseq}). Therefore, we can use the same reasoning to derive the solution to this equation. This leads to the fact that we should ignore the part of the solution involving the modified Bessel function of the first kind, which blows up at the radial infinity. Therefore, the solution of  (\ref{scalarwcseq})  will only involve the modified Bessel function of the second kind, which can be written without loss of generality as,
\begin{equation}
	\varphi(r)=\frac{\kappa' (\tilde\mu+\tilde T)-4c_1'}{2\pi}\, K_0(m r) .
\end{equation}

The solution now involves an arbitrary integration constant, $c_1'$.
To determine $c'_1$, we can use the fact that when $\tilde \mu\rightarrow \tilde T$, the solution should be reduced to the corresponding straight string solution \eqref{strmasspot}. This boundary condition indicates that $c'_1= 0$. Hence, finally, for the massive scalar field, we obtain
\begin{eqnarray}
    \varphi(r)= \frac{\kappa'(\tilde{\mu} +\tilde{T})}{2\pi}K_0(m r).
\end{eqnarray}
The tensorial components $\theta_{\mu\nu}$ do not change and the nonzero metric perturbation tensor components become
\begin{eqnarray}
 	&&h_{00}=-\frac{\kappa (\tilde\mu-\tilde T)}{2\pi G_4(0)}\ln r+\frac{\kappa'(\tilde{\mu} +\tilde{T})}{2\pi}\frac{G_{4,\phi}(0)}{G_4(0)} K_0(m r),\quad  \label{wigglyhoomassive}\\
 	 &&h_{zz}=-\frac{\kappa (\tilde\mu-\tilde T)}{2\pi G_4(0)}\ln r-\frac{\kappa'(\tilde{\mu} +\tilde{T})}{2\pi}\frac{G_{4,\phi}(0)}{G_4(0)}\, K_0(m r),\ \ \
 	\\
 	&&h_{xx}=h_{yy}=-\frac{\kappa (\tilde\mu+\tilde T)}{2\pi G_4(0)}\ln r -\frac{\kappa'(\tilde{\mu} +\tilde{T})}{2\pi}\frac{G_{4,\phi}(0)}{G_4(0)}\, K_0(m r).
\end{eqnarray}

The line element around a wiggly cosmic string in the presence of a massive scalar field in the Horndeski theory can be written as
 \begin{eqnarray}
 	ds^2=\left[1-\frac{\kappa'(\tilde{\mu} +\tilde{T})}{2\pi}\frac{G_{4,\phi}(0)}{G_4(0)} K_0(m r)\right] \Biggl\{ &-&\left[1+\frac{\kappa (\tilde\mu-\tilde T)}{2\pi G_4(0)}\ln r\right] dt^2 + \left[1-\frac{\kappa (\tilde\mu-\tilde T)}{2\pi G_4(0)} \ln r \right] dz^2  \nonumber   \\
 	&+& \left[1-\frac{\kappa \left( \tilde \mu+\tilde T\right)}{2\pi G_4(0)}\ln r\right] \left(dr^2+r^2 d\theta^2\right) \Biggr\}. \label{wigglycsmassive}
 \end{eqnarray} 

This solution is compatible with the literature \cite{VilenkinShellard} in the appropriate GR limit. Similar to the massless scalar field case, this solution can also be put in a more convenient form by applying the transformation similar to (\ref{transf1}), which gives
\begin{eqnarray}\label{wsmassive}
 	ds^2=\left[1-\frac{\kappa'(\tilde{\mu} +\tilde{T})}{2\pi}\frac{G_{4,\phi}(0)}{G_4(0)} K_0(m r')\right] \Bigg\{&-&\left[1+\frac{\kappa (\tilde\mu-\tilde T)}{2\pi G_4(0)}\ln r' \right] dt^2 + \left[1-\frac{\kappa (\tilde\mu-\tilde T)}{2\pi G_4(0)} \ln r' \right] dz^2  \nonumber   \\
 	&+& dr'^2+ \left[1-\frac{\kappa \left( \tilde \mu+\tilde T\right)}{2\pi G_4(0)}\right] r'^2 d\theta^2 \Bigg\}.
\end{eqnarray}

We  again see that the surfaces of constant $t$ and $z$ are conformally equivalent to cones with deficit angle  $\kappa (\tilde \mu +\tilde T)/[2\pi G_4(0)]$. We study further properties of wiggly strings for both cases in the next section.
\begin{figure}[t] 
\includegraphics[draft=false,width=.7\linewidth]{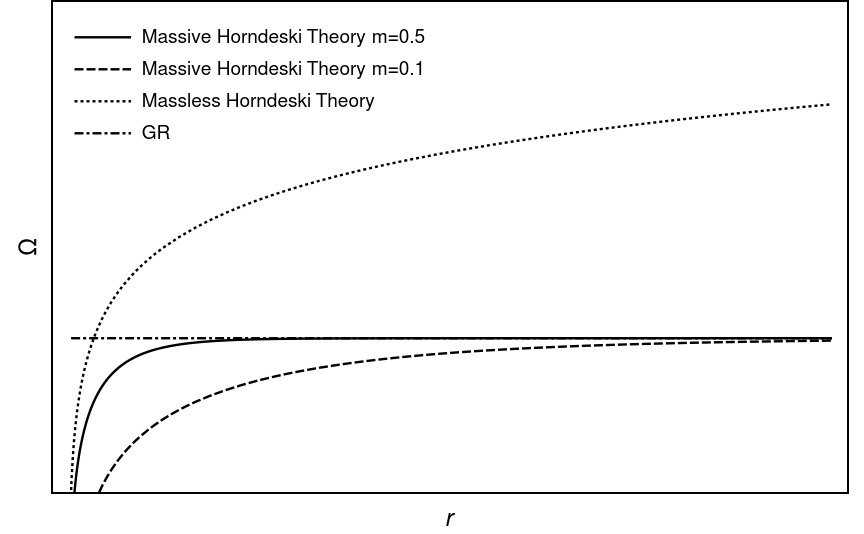}
		\caption{Behavior of the conformal factors for wiggly cosmic strings in linearized Horndeski theory. We observe the same graphical trend as the straight string! Here $\Omega$ represents the conformal factors of wiggly cosmic string solutions of the corresponding cases (\ref{wigglycsmassless}) and (\ref{wigglycsmassive}).\raggedright}
	\label{fig:4}
\end{figure}
\section{Properties of Wiggly Cosmic strings}

\subsection{Conformal factor} 
Let us discuss some properties of wiggly strings in linearized Horndeski theory. The form of the solutions of the Horndeski theory is conformal to wiggly strings in GR. Hence, the conformal factor determines some of their characteristic behavior. For example, for the massless Horndeski theory, the conformal factor increases unboundedly, similar to straight strings in Horndeski theory. Hence, the solution is only valid near the string, and for the far region, an exact solution may be needed to determine the properties of these strings. For the massive Horndeski theory, however, the conformal factor approaches to unity far from the string, and the effect of the scalar field is screened. Therefore, the solution resembles the wiggly string solution in GR, far from the symmetry axis where the string lies. If we plot these conformal factors, we will obtain graphs given in Figure (\ref{fig:4}), which are similar to the graphs of straight strings presented in the Figure (\ref{fig:1}).  Hence, the mass of the scalar field screens the effect of the scalar field in the far region, similar to the behavior of straight strings we have studied in section III or global monopoles in massive Horndeski theory \cite{Secukmonopole} in a recent work.      
 \begin{figure}[t] 
\includegraphics[draft=false,width=.7\linewidth]{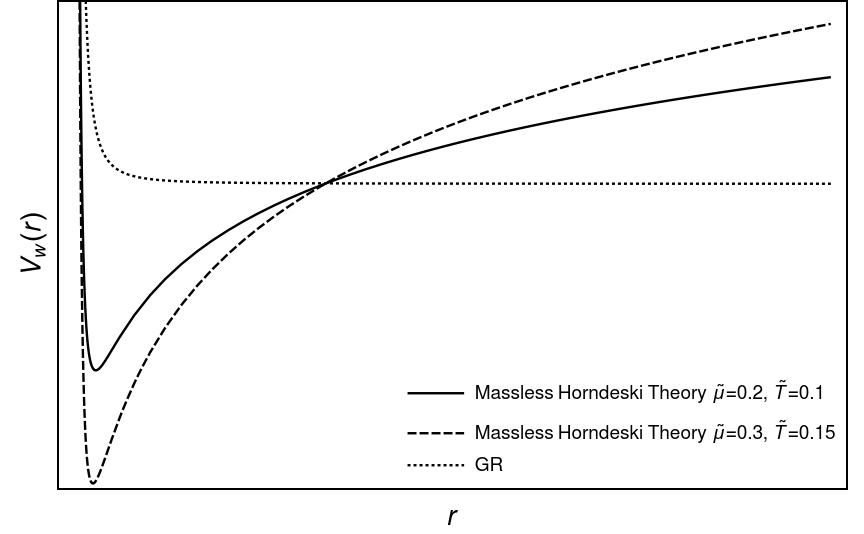}
	\caption{Behavior of the effective potential of massless Horndeski theory with $\tilde\mu=0.2, \tilde T=0.1 $ and $\tilde\mu=0.3, \tilde T=0.15 $. Graph is plotted for $L=0.01$. Other constants are chosen such that the behavior is reflected on a single graph.\raggedright
    }
    \label{fig:5}
\end{figure}
\subsection{Geodesics and effective potential}

We consider motion perpendicular to the symmetry axis of the string. Therefore, we take $dz=0$ in (\ref{wigglycsmassless}) for massless and in (\ref{wigglycsmassive})  for massive Horndeski theory. The following transformation
\begin{equation}\label{wigdztrans}
\left[1-\frac{\kappa \tilde \mu}{\pi G_4(0)}\ln r\right]r^2=\left[1-\frac{\kappa \tilde \mu}{\pi G_4(0)}\right] r''^2
\end{equation}
brings the corresponding line elements (\ref{wigglycsmassless}) or (\ref{wigglycsmassive}) of the form
\begin{align}\label{wiggly3d}
 ds^2 = A_w(r)\left[ -dt^2+dr^2+\left(1-\beta\right) r^2 d\theta^2 \right]
 \end{align}
where we  have removed double prime for clarity, $\beta$ represents the angular deficit parameter in these coordinates, given by
\begin{equation}\label{beta}
     \beta = \frac{\kappa \tilde \mu}{\pi\, G_4(0)},
\end{equation}
and $A_w$ describes the conformal factors of the wiggly cosmic string solutions after the transformation (\ref{wigdztrans}). For the massless case we have the following conformal factor 
\begin{align}\label{wsmasslessconf}
 A_w(r) = \left[1 + \frac{\kappa (\tilde\mu-\tilde T)}{2\pi G_4(0)}\ln r+
                \frac{G_{4,\phi}(0)\kappa' (\tilde\mu+\tilde T)}{2\pi\,G_4(0)}\ln r
            \right], 
\end{align}
and for massive case the conformal factor now becomes 
\begin{align}\label{wsmassiveconf}
 A_w(r) = \left[1 + \frac{\kappa (\tilde\mu-\tilde T)}{2\pi G_4(0)}\ln r
                -\frac{G_{4,\phi}(0)\kappa'(\tilde{\mu} +\tilde{T})}{2\pi G_4(0)} K_0(m r)
            \right]&.
\end{align}

The line  element (\ref{wiggly3d}), given for $dz=0$, resembles again the one given in \cite{VilenkinShellard}. For the subspace perpendicular to the symmetry axis of the string, the line element is conformal to a straight string solution of GR, with the differences being those of the mass per unit length and coupling constant. The conformal factor clearly differs from straight strings in Horndeski theory.

\begin{figure}[t]
  \begin{subfigure}[t]{0.5\linewidth}
    \centering
    \includegraphics[draft=false,width=.91\linewidth]{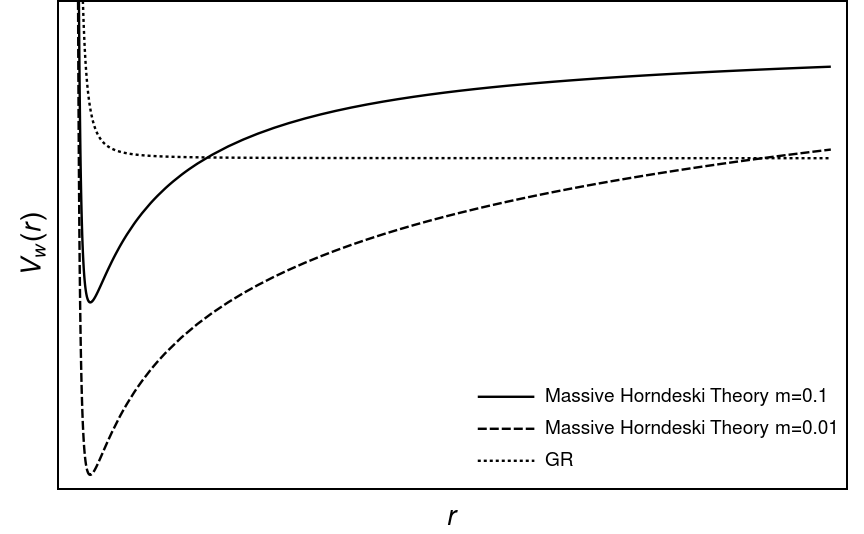} 
    \caption{Behavior of the effective potential of massive Horndeski theory with $m=0.1$ and $m=0.01$. \raggedright
    }
    \label{fig6:a} 
    \vspace{2ex}
  \end{subfigure}
  \begin{subfigure}[t]{0.5\linewidth}
    \centering
    \includegraphics[draft=false,width=.91\linewidth]{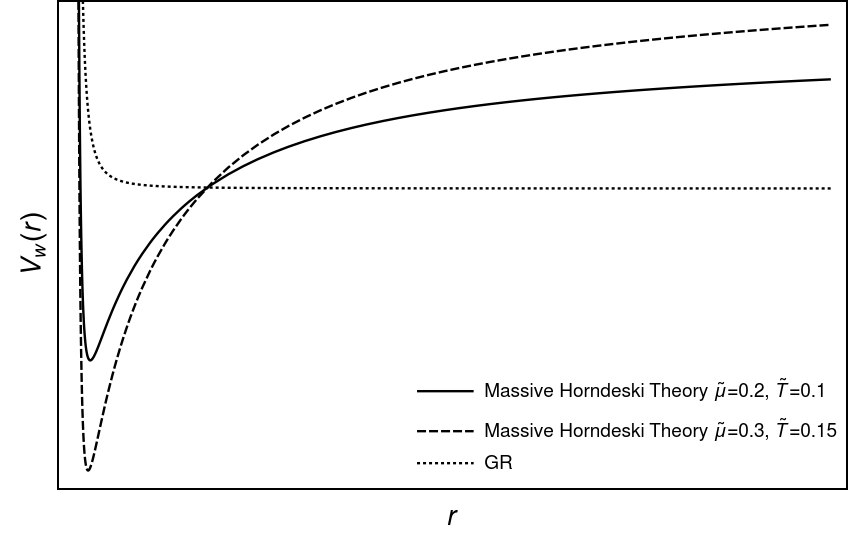} 
    \caption{Behavior of the effective potential of massive Horndeski theory with $\tilde\mu=0.2, \tilde T=0.15 $ and $\tilde\mu=0.3, \tilde T=0.15 $.\raggedright
    } 
    \label{fig6:b} 
  \end{subfigure}
    \caption{Contrasted to straight strings in massive Horndeski theory, the effective potential for wiggly strings does not converge. Therefore, we do not obtain the asymptotic behavior of general relativity, since the conformal factor $A_w(r)$ is dominated by the logarithmic function. Both graphs are plotted for $L=0.1$ and have the same range. In (\ref{fig6:a}), $\tilde\mu =0.2$, $\tilde T=0.1$ and in (\ref{fig6:b}), $m=0.2$. Note that plots are not scaled. We choose constant values such that these behaviors are clear to see on a single graph.\raggedright
    }\label{fig:6}
\end{figure}

Since the form of the line element (\ref{wiggly3d}) resembles that of straight strings  (\ref{cshornmetric}) in massless or massive Horndeski theory with the identification $A(r)\rightarrow A_w(r)$, $\mu\rightarrow\tilde \mu$, and $\alpha\rightarrow \beta$, we can use the general equations of motion of particles around straight strings  presented in section (\ref{section4c}) also for  wiggly strings as well for the motion of test particles moving in the plane perpendicular to the symmetry axis of the string defined by $\dot{z}=0$. Note that, in general, the results may differ, since conformal factors $A_w$ are different for wiggly strings compared to straight ones for both cases. Coincidentally, we obtain the general equation of effective potential for wiggly cosmic strings, similar to straight ones (\ref{effpot}) in linear Horndeski theory, as follows:
\begin{equation}\label{Veffwg}
V(r)=\frac{L^2}{(1-\beta)r^2}-\epsilon A_w(r).
\end{equation}
We observe that in Fig. (\ref{fig:5}), the general trend of the effective potential of the wiggly strings in massless Horndeski theory does not change, as expected, compared to straight strings. Since the leading function of the effective potential in massless theory for both straight and wiggly strings is proportional to $\ln(r)$. If we consider the massive Horndeski theory, however, we see that the general behavior of the effective potential has important differences between straight and wiggly strings. As shown in Fig. (\ref{fig:6}), we see that the potential is no longer bounded at radial infinity, since now the conformal factor is dominated by the logarithmic function as opposed to straight strings in massive theory where the conformal factor is bounded from above and tend to reach GR result, as clearly seen in Fig.  (\ref{fig:3}). Hence, at infinity, we observe a similar behavior of the effective potential for massless or massive Horndeski theories for wiggly strings.

\subsection{Circular geodesics}

The shape of the effective potential (\ref{Veffwg}) for wiggly strings implies that, similar to straight ones, there are circular geodesics around these strings for both massive and massless Horndeski theory. The general equations for circular geodesics have the same form as straight strings, as given in (\ref{E},\ref{L},\ref{Vpp}), with the identifications $A\rightarrow A_w$, $\alpha \rightarrow \beta$. For the massless Horndeski theory, the specific energy and angular momentum of a test article following a circular geodesic around a wiggly cosmic string are given by:
\begin{eqnarray}
E^2&=&1+\frac{L^2}{(1-\beta)r_c^2}+
\frac{\kappa (\tilde\mu-\tilde T)}{\pi\, G_4(0)}\ln r_c+
                \frac{G_{4,\phi}(0)\kappa' (\tilde\mu+\tilde T)}{\pi\,G_4(0)}\ln r_c,
\\
L^2&=&\frac{(1-\beta)}{4\pi\, G_{4}(0)}\left[\kappa (\tilde \mu-\tilde T)+G_{4,\phi}(0) \kappa'(\tilde \mu+\tilde T)\,  \right]r_c^2.
\end{eqnarray}
These circular geodesics are stable if $V''>0$, which is
\begin{equation}
    V''(r_c)=\frac{1}{\pi\, G_4(0)\, r_c^2}\left[\kappa(\tilde \mu-\tilde T)+ G_{4,\phi}(0)\kappa' (\tilde \mu +\tilde T)\right].
\end{equation}
Hence the circular geodesics are definitely stable for physically favorable case $\tilde \mu> \tilde T$.

An important difference between straight and wiggly strings, shown in these expressions, is that for wiggly strings, even in the GR limit ($\kappa'=0$), circular geodesics exist. These expressions also show that the specific energy and angular momentum should be greater for a particle with the same radius to follow a circular geodesic compared to straight strings for $\tilde \mu> \tilde T$ or vice versa.

The existence of circular geodesics is another indication of the fact that the wiggly strings are gravitationally active and accumulate matter around them, similar to the straight strings in Horndeski theory. Note that the wiggly strings still have this behavior in the GR limit, as opposed to the straight ones. 

\section{Velocity Kick on particles passing through a string in (massive) Horndeski theory }

In the sections IV and VI we have discussed  bounded geodesics around straight and wiggly strings. Here, we will consider unbounded motion of particles passing through such strings. 
Hence, we will investigate the effect of wiggly and straight strings in such trajectories. We will assume large velocity for the particle and small impact parameter such that $v^2>>h_{00}$ is satisfied \cite{VilenkinShellard}. We start with the line element of the form \eqref{wiggly3d} in the Cartesian coordinates as,
\begin{equation}
ds^2 = f(x,y)\left[ -dt^2 + dx^2 + dy^2 \right],
\end{equation}
with a angular deficit $\tilde{\Delta}=\pi\beta$ where $\beta$ is given in ($\ref{beta}$) (for regular cosmic strings, $\tilde\Delta=\pi\alpha$ where $\alpha$ is given in (\ref{alpha})). Here, $f(x,y)$ represents the conformal factor of various cosmic string solutions we have derived in this paper in the Cartesian coordinates. Note that this function takes distinct forms with respect to whether the string is wiggly or straight and/or whether the scalar field is massive or massless. The linearized equations of motion of the particle immediately follow as,
\begin{align}
    \partial_t^2 x = -\frac{1}{2}\left[1 - (\partial_t x)^2 - (\partial_t y)^2\right]\partial_x f(x,y),
    \\
    \partial_t^2 y = -\frac{1}{2}\left[1 - (\partial_t x)^2 - (\partial_t y)^2\right]\partial_y f(x,y).
\end{align}
Following \cite{VilenkinShellard}, we choose the $x$-axis as the initial trajectory with conditions $x=v\,t, y=y_0$. Then, after passing the string, the small velocity gain of the particle in its $y$-component can be found as
\begin{equation}\label{mainintegral}
    v_y = \frac{1}{v} \int_{-\infty}^{\infty} \partial_t^2 y \, dx
        = -\frac{1}{2v\gamma^2} \int_{-\infty}^{\infty} \partial_y f(x,y)\bigg|_{y=y_0} dx,
\end{equation}
where $\gamma^2=1/(1-v^2)$ and $y\in\mathbb{R}$. In the following, we will solve this integral separately for all cases we investigated throughout the paper.

\subsubsection{Case I: Straight String in Massless Horndeski Theory}
For straight strings in massless Horndeski theory the conformal factor can be read from \eqref{csmassless},
\begin{equation}\label{smasslessconf}
f(x,y)= 1+\frac{G_{4,\phi}(0)}{\pi G_4(0)}\kappa'\mu \, \ln\sqrt{x^2+y^2}.\nonumber
\end{equation}
Therefore the trajectory integral \eqref{mainintegral} is,
\begin{align}
    v_y&= -\frac{\kappa'\mu\,G_{4,\phi}(0)}{2\pi v\gamma^2G_4(0)}
          \int_{-\infty}^{\infty} \partial_y\ln \sqrt{x^2+y^2}\bigg|_{y=y_0}dx, \nonumber
          \\
       &=-\text{Sgn}(y_0)\,\frac{\kappa'\mu\,G_{4,\phi}(0)}{2v\,\gamma^2\,G_4(0)}.
\end{align}
The resulting sign changes by $y_0$ being negative or positive. Now imagine two particles with the same initial velocity $v$ passing through the opposite side of the string. Meaning that one particle propagates with $y_0>0$, while the other $y_0<0$. Hence, the velocity difference of these particles is given as,
\begin{equation}\label{speeddiffmasslessstraigt}
    |\Delta v_y|=\frac{\mu}{G_4(0)} \left[v\kappa
    +\frac{\kappa'\,G_{4,\phi}(0)}{v\,\gamma^2}\right].
\end{equation}
Here, the first term comes from the deficit angle $\tilde\Delta=\pi\alpha$. Note that for straight cosmic strings in GR, there is no gravitational force exerted from the source. The change in speed in $y$-direction represented by the first term in \eqref{speeddiffmasslessstraigt} is due to the global structure of the string topology \cite{VilenkinShellard}. The second term is the contribution of the massless Horndeski theory, where the straight strings apply a gravitational force on the particles around them. With this term, we observe that the velocity gain is a boost relative to GR, since $v,\kappa',\gamma>0$.

\subsubsection{Case II: Straight String in Massive Horndeski Theory}
In this case the conformal factor given as \eqref{csmassive},
\begin{equation}
f(x,y)=1-\frac{ G_{4,\phi}(0)}{\pi G_4(0)}\kappa'\mu\, K_0\left( m \sqrt{x^2+y^2} \right). \nonumber
\end{equation}
The velocity integral \eqref{mainintegral} is,
\begin{align}\label{speedy}
    v_y&= \frac{ G_{4,\phi}(0)}{2\pi v\gamma^2 G_4(0)}\kappa'\mu
          \int_{-\infty}^{\infty} \partial_y K_0 \left( m \sqrt{x^2+y^2}\right)\bigg|_{y=y_0}dx, \nonumber
          \\
       &=-\text{Sgn}(y_0)\frac{ G_{4,\phi}(0)\kappa'\mu}{2v\gamma^2G_4(0)} e^{-m|y_0|}\ .
\end{align}
For the solutions of the integrals involving Bessel functions, see \cite{Watson}. The velocity difference in this case is,
\begin{equation}
     |\Delta v_y|=\frac{\mu}{G_4(0)} 
                  \left(v\kappa + \frac{\kappa'\,G_{4,\phi}(0)}{v\,\gamma^2}e^{-m|y_0|}\right).
\end{equation}
Comparing this result to our previous one, we see that, in the massive Horndeski theory, the initial placement of the particle in the $y$-axis directly affects the speed of the particle, $v_y$ due to the coupling of the mass of the scalar field, $m$ and the impact parameter $y_0$ in $e^{-m |y_0|}$. The effect of this term diminishes as $y_0\rightarrow\infty$, and the result approaches to GR one. This is the result of the screening effect of the massive scalar field, which we have also observed in the effective potential analysis in the earlier sections. Furthermore, we recover the velocity difference expression for straight strings  (\ref{speeddiffmasslessstraigt}), when $m\rightarrow 0$.

\subsubsection{Case III: Wiggly String in Massless Horndeski Theory}

In the previous two cases, we have mainly calculated all the necessary integrals. Hence, in this case and the next one, we will present our results directly. The conformal factor for massless Horndeski theory is given explicitly as \eqref{wsmasslessconf}, and in Cartesian coordinates it reads,
\begin{equation}
f(x,y) = 1 + \frac{\ln\sqrt{x^2+y^2}}{2\pi\, G_4(0)}\left[\kappa(\tilde\mu-\tilde T) +\kappa'(\tilde\mu+\tilde T)\,G_{4,\phi}(0)\right].\nonumber
\end{equation}
Since the leading function is exactly the same as massless straight strings \eqref{smasslessconf}, the result of velocity gain is immediate,
\begin{align}
    v_y = -\text{Sgn}(y_0)\frac{\kappa(\tilde\mu-\tilde T)+\kappa'(\tilde\mu+\tilde T)G_{4,\phi}(0)}{4v\gamma^2G_4(0)}.
\end{align}
The velocity difference is found as,
\begin{equation}
    |\Delta v_y|=\frac{\tilde\mu v\kappa}{G_4(0)} + \frac{1}{2v\gamma^2G_4(0)}\left[
    \kappa(\tilde\mu-\tilde T) + \kappa'(\tilde\mu+\tilde T) G_{4,\phi}(0)\right].
\end{equation}
This expression differs from our straight string result \eqref{speeddiffmasslessstraigt}, with the addition of the second term due to $\tilde \mu \neq \tilde T$ for wiggly strings, which is the consequence of the gravitational force of the wiggly string, in which this term is also present in GR \cite{VilenkinShellard}. The last term is due to the scalar field of Horndeski theory for a wiggly string, which is not present in the corresponding GR expression \cite{VilenkinShellard}. 

\subsubsection{Case IV: Wiggly String in Massive Horndeski Theory}

Using the conformal factor \eqref{wsmassiveconf} in Cartesian coordinates, we found the velocity gain for massive Horndeski theory for wiggly strings as,
\begin{align}
    v_y = -\frac{\text{Sgn}(y_0)}{4v\,\gamma^2G_4(0)}
    \left[
    \kappa (\tilde\mu-\tilde T)+\kappa'(\tilde{\mu} +\tilde{T})G_{4,\phi}(0)e^{-m|y_0|}
    \right],
\end{align}
and the velocity difference,
\begin{equation}
    |\Delta v_y|= \frac{\tilde\mu v\kappa}{G_4(0)}
    +\frac{1}{2v\,\gamma^2G_4(0)}
    \left[
    \kappa (\tilde\mu-\tilde T)+\kappa'(\tilde{\mu} +\tilde{T})G_{4,\phi}(0)e^{-m|y_0|}
    \right].
\end{equation}
From this general expression, one can deduce the three previous cases of $v_y$. For example, to obtain our massless straight string result, one puts $m\rightarrow0, \tilde\mu=\tilde T=\mu$ and obtains Eq. \eqref{speeddiffmasslessstraigt}. The screening effect again reveals itself in the last term, where the effect diminishes with increasing $|y_0|$ and disappears as $|y_0|\rightarrow \infty$, far from the string, reducing the expression to the GR one. 

In this section, we derived the change in the velocity of particles passing through a cosmic string in various settings. We have obtained the corresponding velocity gain of the particles, which basically comes from three effects. The first one is the topological term due to the global conicity of the spacetime, which is present for all cases. The second one is due to the gravitational force of the strings, which does not exist only for straight strings in GR. The last term is the effect of the scalar field, nonminimally coupled to gravity. Hence, it  exists as long as  $G_{4,\phi}\neq 0$. In other words, this effect is not present for the theories with $G_{4,\phi}= 0$ such as GR, Einstein-scalar theory, Quintessence/K-Essence, etc, where either there is no scalar field or the coupling of the scalar field to the gravity is minimal. The last term also carries the effects of the mass of the scalar field as a decay factor, with the screening effect, which only exists for the massive scalar theories, such as BD with a potential, $f(R)$ theories, etc. Note that these results are valid as long as the linear approximation is valid.  

In astrophysical or cosmological applications, this expression can also easily be used to calculate the relative velocity, that is, $u=\gamma|\tilde\Delta v_y|$, where the effects of  a string relativistically moving in a perfect fluid background can be considered. The imprints of the wakes due to a relativistic string can be investigated in a dark matter background or in the CMB, and observational consequences can be studied, using the results obtained here for the strings of the massless or massive Horndeski theory, in future works.  

 \section{Conclusion}
In this paper, we first discussed solutions of linearized field equations of Horndeski theory corresponding to a static straight cosmic string lying along the $z$ axis  for both massless and massive scalar field cases. We have obtained the spacetime line element and the scalar field configuration for both cases in appropriate forms, which makes the comparison easier to the known versions of  corresponding solutions for GR and BD theories. The solutions of  massless and massive theories  have distinct features, which can be tractable in the behavior of the conformal factors of the corresponding line elements and test particle motion around the strings.  The existence of the mass of the scalar field results in a screening effect, and the scalar field effectively freezes from a distance from the string. The result of this behavior is that the cosmic strings in massive theory are asymptotically indistinguishable from corresponding ones in GR, except for the differences in the coupling constants in the angular deficit parameter. For the massless theory, however, the behavior of the strings is very different. The conformal factor asymptotically  increases, undoubtedly making the solution valid only in regions not very far from the string. Moreover, the particles make a bounded motion, and stable circular geodesics exist for a particle around the string at constant $z$.  

The second type of solutions we have presented are the solutions corresponding to wiggly cosmic strings for both massless and massive theory. We have again obtained the solutions for both cases and presented them in appropriate forms for further analysis. The solutions resemble the form of  the earlier solutions found for GR and BD, where the BD solution is conformal to GR. The difference between massless and massive theories is, again, revealed in the conformal factors. The behavior of the conformal factors is just the same as straight strings discussed in sections III and IV in the present manuscript. The mass of the scalar field again shows a screening effect, and wiggly strings in massive Horndeski theory become equivalent to strings in GR when viewed at asymptotically radial distances. Similar to straight  strings, the particles around these strings make bounded motion as long as the weak field approximation is valid.

In VII, we investigated the effects of the straight or wiggly strings in massless or massive Horndeski theory on the particle velocity passing through the string. We generalize the results of \cite{VilenkinShellard} by finding the modifications not only for the effect of the straight string topology and the gravitational force of the wiggly string on the particle velocity, but also for the massive and massless scalar field. We observed velocity gains in both massive and massless linear Horndeski theory for straight and wiggly string space-times.
These velocity "kicks" can have significant effects on the background matter distributions on intergalactic dark matter and cosmic fluid distributions. A future direction stemming from this work is to investigate the motion of a cosmic string in the dark matter and cosmic fluid distributions, revealing the effects of the scalar field and its mass on these fluids by considering the effects of the modifications we have obtained in the Horndeski theory on the cosmic string wakes.    
 
 \section{Acknowledgments}
M. H. S. and O. D. are supported by Marmara University Scientific Research Projects Committee (Project Code: FDK-2021-10432).

\appendix

\section{ Linearized field equations of Horndeski Theory \label{Appendix}}

We consider a weak field expansion of the metric tensor and the scalar field as
\begin{eqnarray}
	&&	g_{\mu\nu}=\eta_{\mu\nu} + h_{\mu\nu},\\
	&&	\phi=\phi_0+ \varphi,	
\end{eqnarray}
with  $\eta_{\mu\nu}$ is the Minkowski metric,  $h_{\mu\nu}$ is the metric perturbation tensor and  $\phi_0$ a constant. Any arbitrary function of the scalar field and the kinetic term $X$ can also be expanded as
\begin{equation}
	f(\phi,X)=f(\phi_0,X_0)+f_{,\phi}(\phi_0,X_0)(\phi-\phi_0)+f_{,X}(\phi_0,X_0)(X-X_0)+\ldots \approx f{(\phi_0,0)}+f_{,\phi}(\phi_0,0)\varphi
\end{equation}
around $\phi=\phi_0$ and $X=X_0=0$. Hereafter we use the abbreviation  $f(\phi_0,0)=f(0)$ for clarity.

Reading up linear terms in \cite{Kobayashi} and adding the matter-energy Lagrangian and energy-momentum tensor, we have the following field equations in linear order in $h$ and $\varphi$ as
\begin{eqnarray}
	&&-\frac{1}{2}K(0)\eta_{\mu\nu}+G_4(0) G_{\mu\nu}^{(1)}+\eta_{\mu\nu}G_{4,\phi}(0) \square_\eta {\varphi}-G_{4,\phi}(0) \varphi_{,\mu\nu} =\kappa T_{\mu\nu}^{(1)},\label{fe1}
	\\
	&&K,_\phi(0)+K,_{\phi\phi}(0)\varphi+\left[K_{,X}(0)-2 G_{3,\phi}(0)\right]\square_\eta {\varphi}+G_{4,\phi}(0)R^{(1)}=0.	\label{fe2}
\end{eqnarray}
We want to explore the case where we have a Minkowski background and hence we set $K(0)=K_{,\phi}(0)=0$. We also use the equations (\ref{fe1}) into (\ref{fe2}) which brings the scalar field equation as
\begin{equation}
	(\square_\eta -m^2)\varphi=\kappa'\, T^{(1)},\label{scalarheq}
\end{equation}
where $ T^{(1)}$ is the trace of the energy-momentum tensor  with 
\begin{eqnarray}
	m^2=-\frac{K_{,\phi\phi}(0)}{K_{,X}(0)-2G_{3,\phi}(0)+3G_{4,\phi}^2/G_4(0)}, \label{massterm}\\
	\kappa'=\frac{G_{4,\phi}(0)\kappa}{G_{4}(0)\left[K_{,X}(0)-2 G_{3,\phi}(0)\right]+3 G_{4,\phi}^2(0) }. \label{kappaprime}
\end{eqnarray}
The equation (\ref{fe1}) can also be put in a convenient form by expanding $G_{\mu\nu}^{(1)}$ in terms of $h_{\mu\nu}$ terms and then transforming the metric perturbation terms into  a new perturbation tensor $\theta_{\mu\nu}$ defined as
\begin{equation}
	\theta_{\mu\nu }\equiv h_{\mu\nu}-\frac{1}{2}\eta_{\mu\nu}h-\eta_{\mu\nu}\frac{G_{4,\phi}(0)}{G_4(0)}\varphi. \label{thetadef}
\end{equation}  
We also consider a Lorenz gauge as 
\begin{equation}\label{Lorenz}
	\theta^\mu_{\phantom{\mu}\nu,\mu}=0.
\end{equation}
Then, the equation (\ref{fe1}) takes a much simpler form as
\begin{equation}
	\square_\eta \theta_{\mu\nu}=-\frac{ 2\kappa}{G_4(0)}\, T_{\mu\nu}^{(1)}. \label{tensorheq}
\end{equation}

The equations (\ref{tensorheq}) and (\ref{scalarheq}), together with the gauge condition (\ref{Lorenz}) are the main equations we will use to obtain the desired linearized solutions.

\end{document}